\documentclass[aps,pra,twocolumn,english]{revtex4-2}
\usepackage[utf8]{inputenc}
\usepackage[T1]{fontenc}
\usepackage{graphicx,lipsum}% Include figure files
\usepackage{dcolumn}% Align table columns on decimal point
\usepackage{bm}% bold math
\usepackage[dvipsnames]{xcolor}
\usepackage{array,multirow}
\usepackage{bm}% bold math
\usepackage{bbm}
\usepackage{amsmath}
\usepackage{ragged2e}
\usepackage[normalem]{ulem} 
\usepackage{mathrsfs}
\usepackage{graphicx}

\begin{document}

% \preprint{APS/123-QED}
%\title {Maximizing Work Extraction from Controlled Quantum Systems Using Pulse Sequences }
\title {Work Extraction from a Controlled Quantum Emitter}


\author{ Kavalambramalil George Paulson, Hanna Terletska}
\affiliation{Department of Physics and Astronomy, Middle Tennessee State University, Murfreesboro, TN 37132, USA}
\email{ paulson.kavalambramalilgeorge@mtsu.edu}

\author{Herbert F Fotso}
\affiliation{Department of Physics, University at Buffalo SUNY, Buffalo, New York 14260, USA}

% \altaffiliation[]{}%Lines break automatically or can be forced with \\
% \author{ }
% \email{}
% \author{ }
% \email{ }
% %\author{\textsuperscript{c}}
% % \email{subhashish@iitj.ac.in}
% \affiliation{}

%\textsuperscript{b,c}}
%\affiliation{
 %Third institution, the second for Charlie Author
%\affiliation{%


\date{\today}% It is always \today, today,

\begin{abstract}
We investigate how an external driving field can control the amount of extractable work from a quantum emitter, a two-level quantum system (TLS) interacting with a photonic environment. In this scenario, the TLS functions as a quantum battery, interacting with the photonic bath that discharges it while the control field recharges it. Ergotropy serves as our measure of the extractable work from the quantum system. We systematically analyze how the ergotropy of the system evolves as it interacts with the photonic bath under the control of either a continuous driving field or a periodic pulse sequence. The coherent and incoherent contributions to the total ergotropy for various initial states are calculated. The role of detuning between the driving field and the emission frequency of the TLS, as well as the initial state of the system in work extraction, are investigated for continuous and periodic pulse-driving fields.
%We use two control strategies with periodic pulse-driven control fields, where the TLS is subjected to near-resonant Rabi driving fields.  The first strategy involves an instantaneous application of the pulse, each pulse implements $\pi$ rad rotation to charge the TLS. In the second strategy, we switch on the $\pi$ pulse for $t_{\pi}$ duration, charging the battery over a set time interval.
We show that detuning has little impact on work extraction for a system that is driven by a periodic sequence of instantaneous pulses. However, for a continuously driven system, as the system approaches its steady state, ergotropy increases with detuning increases. %, especially in the regime of near-steady states' dynamics.

%an instantaneous pulse-driven system, whereas, for a continuous-driven system,  ergotropy increases as detuning increases, especially in the regime of near-steady states' dynamics.
%For the $\pi$ pulse-driven system, work extraction dynamics are consistent with the continuous-driven system for shorter inter-pulse delays.
%\begin{description
%\end{description}
\end{abstract}

\maketitle

%\tableofcontents

\section{\label{sec:level1} Introduction}
The ability to precisely control and manipulate quantum systems is paramount in quantum information processing and computation. The delicate nature of quantum states necessitates sophisticated strategies to achieve desired outcomes, making the control and tuning of quantum systems a focal point of research.% and development.
%As it is known a closed quantum system lacks generality in its nature and the interaction between a system and its surrounding environment facilitates the exchange of both matter and energy, thereby inducing non-trivial alterations in the dynamics of the quantum system, {\color{red} leading to decoherence and dissipation of the quantum state}. 

Ideally, quantum information processing platforms should be isolated from their environment. However, ideal closed quantum systems are rare in nature. Most quantum systems interact with their surrounding environment, facilitating the exchange of both matter and energy. These interactions induce significant and complex alterations in the dynamics of the quantum system, often leading to decoherence and dissipation of the quantum state. This highlights the necessity of effective control techniques to regulate the environment's influence on quantum systems. Thus, successfully implementing diverse quantum architectures relies on adeptly managing this environmental impact. Various approaches have been developed to mitigate the decoherence and dissipation arising from the system's interaction with its environment~\cite{viola1998dynamical,shiokawa2004dynamical,kuopanportti2008suppression}. In the case of photon-mediated operations, manipulating the absorption and emission of a quantum system and forcing it to occur at specific frequencies is a pivotal aspect of quantum control, achieved through exposure to either a continuous laser field or a sequence of optical pulses~\cite{patel2010two,gao2015coherent, fotso2016suppressing,fotso2017absorption,fotso2018controlling,lukin2020spectrally}. Such protocols have been shown to have potentially dramatic effects on the performance of photon-mediated quantum information processing operations ~\cite{fotsoTPIprb2019,lukinNPJ2020,Joas2017}, making it important to investigate the impact of such control protocols on the other properties of the quantum emitters. 

%when an external field is applied to the quantum system, its population dynamics change, and so does the energy. this makes comprehending population dynamics quite significant in the field of quantum information processing and computation. \newline

% Comprehending the population dynamics of a quantum system becomes crucial when an external field is applied, as it alters the system's energy and population distribution, making it a vital aspect of quantum information processing and computation.\newline
% Thermodynamics traditionally explores phenomena like heat, work, temperature, and their interrelations within macroscopic systems. Quantum thermodynamics extends these fundamental concepts to the microscopic scale, encompassing quantum systems' characteristics and dynamics.\newline
% \textcolor{blue}{In the field of quantum thermodynamics, quantum batteries have emerged as a prominent subject of study. These batteries are quantum systems capable of releasing a finite amount of energy as work through unitary cyclic processes. The literature on quantum batteries predominantly focuses on two aspects: the process of charging them and the extraction of work from them.}

When a control field is applied to a quantum system, it injects additional energy into the system, consequently modifying its population dynamics and average energy. %\sout{The nonunitary dynamics of dissipation and decoherence, coupled with the control pulse, resembles the process of charging and discharging a quantum battery.} {\color{red} 
The non-unitary dynamics of dissipation and decoherence, coupled with the control pulse, is conceptually analogous to the processes involved in charging and discharging a quantum battery~\cite{alicki2013entanglement, hovhannisyan2013entanglement, perarnau2015extractable,giorgi2015correlation}. A quantum battery, functioning as a quantum thermodynamic system, stores energy at energy levels, and coherence is central to energy storage and utilization. 
Significant research has focused on developing various quantum battery models ~\cite{ferraro2018high,santos2019stable,arjmandi2022enhancing, hu2022optimal,salvia2023quantum} and evaluating their effectiveness. 
%Extensive research has concentrated on crafting diverse quantum battery models
 %and scrutinizing their efficacy. 
 Various methodologies have been employed to maximize the extraction of work from these batteries, aiming to optimize their performance~\cite{gherardini2020stabilizing, mitchison2021charging, yao2022optimal, mazzoncini2023optimal}.

Quantum control methodologies represent potentially powerful tools for unlocking the full capabilities of quantum batteries. By precisely regulating the charging and discharging mechanisms through dynamic modification of control parameters such as pulse duration and frequency, we aim to optimize work extraction and enhance the overall performance of these systems.
%Quantum control methodologies are crucial in unlocking the complete capabilities of quantum batteries. This prompts the intriguing question of the maximum amount of work that can be extracted from the system. We can precisely regulate these batteries' charging and discharging mechanisms by dynamically modifying control parameters like pulse duration and frequency. 
%Such a meticulous control technique improves the effectiveness of energy storage and retrieval and offers opportunities to maximize battery performance across different scenarios. 
Hence, understanding the complexities of quantum control protocols on two-level systems is of great importance from the perspective of using them as quantum batteries. %forward and unleashing their potential applications in quantum technologies and beyond. 
%In the present work, we show that detuning $\Delta$ does not influence the work extraction from a TLS for instantaneous pulse applied, whereas a change in detuning affects the work extraction for continuously driven and a periodic $\pi$  pulse applied. 
%{\color{red} Here we contrast continuously driven and pulse systems. We demonstrate a dramatic effect of $\delta$  on continuous and pulse-driven systems}.

To explore the potential of quantum control methodologies in optimizing quantum batteries, we investigate a two-level system (TLS) interacting with a photonic bath and subjected to external driving fields. Our work examines the potential for work extraction from this non-unitarily evolving system under different conditions and control strategies. We consider various initial states and employ continuous and periodic pulse control methods for charging them. %These quantum control strategies act as switches to charge the batteries. 
In the first protocol, the system is continuously driven by a strong driving field, while the second protocol involves applying a periodic pulse sequence. In both scenarios, we examine how the amount of extractable work and ergotropy from the system varies and how it is connected to the population dynamics of the TLS. 
In this analysis, we primarily consider two initial states for the quantum battery: the excited state with zero coherence (incoherent state) and the maximally coherent state. (In) coherent contributions to the total ergotropy are calculated for both initial states of the quantum battery. We also examine how the detuning between the driving field frequency and the TLS natural emission frequency affects the calculated ergotropy under the influence of either the continuous field or the periodic pulse sequence drive.
%We demonstrate that the efficiency of work extraction can be tuned by varying the timing and duration of pulse applications. 

%We consider various initial quantum battery states and employ continuous and periodic pulse control methods for charging them. These quantum control strategies act as switches to charge the batteries. In the first technique, pulses are applied instantaneously in periodic sequences, enabling rapid charging. The second protocol involves applying pulses for a specific duration. In both scenarios, alongside continuous driving field cases, we examine how the amount of extractable work from the system varies and its connection with the population dynamics of the TLS. We primarily consider two initial states for the quantum battery: excited state with zero coherence and maximally coherent state. (In) coherent contributions to the total ergotropy are calculated for both initial states of the quantum battery. We demonstrate that the efficiency of work extraction can be tuned by varying the timing and duration of pulse applications. Additionally, we examine how detunings affect ergotropy calculations under the influence of both continuous and periodic pulse sequences.      

%This work examines the scenario involving a two-level system interacting with a photonic bath and subjected to external driving fields. We investigate the potential for extracting work from the nonunitarily evolving system. 

This paper is organized as follows: 
The model of the system of a TLS interacting with a photonic bath driven by the control field is presented in Sec.~\ref{model}.
In Sec.~\ref{ergotropy}, we present the theoretical formalism, including a discussion of ergotropy to quantify the amount of extractable work from the system. This section also presents the calculations of both the coherent and the incoherent contributions to the ergotropy.
In Sec.~\ref{erg_TLS}, we present the results for different driving protocols used for charging the quantum battery. We estimate the ergotropy of various quantum states under different driving pulses and establish its connection with population dynamics. Finally, we conclude our findings in Sec.~\ref{conclusion}.

%%%%%%%%%%%%%%%%%%%%%%%%%%%%%%%%%%%%%%%%%%%%%%%%%%%%%%%%%%%%%%%%%%%%%%%%%%%%%%%
\section{Model of a two-level system coupled to the photonic bath}\label{model}
\begin{figure}[t!]
    \justifying
    \includegraphics[height=50mm,width=1\columnwidth]{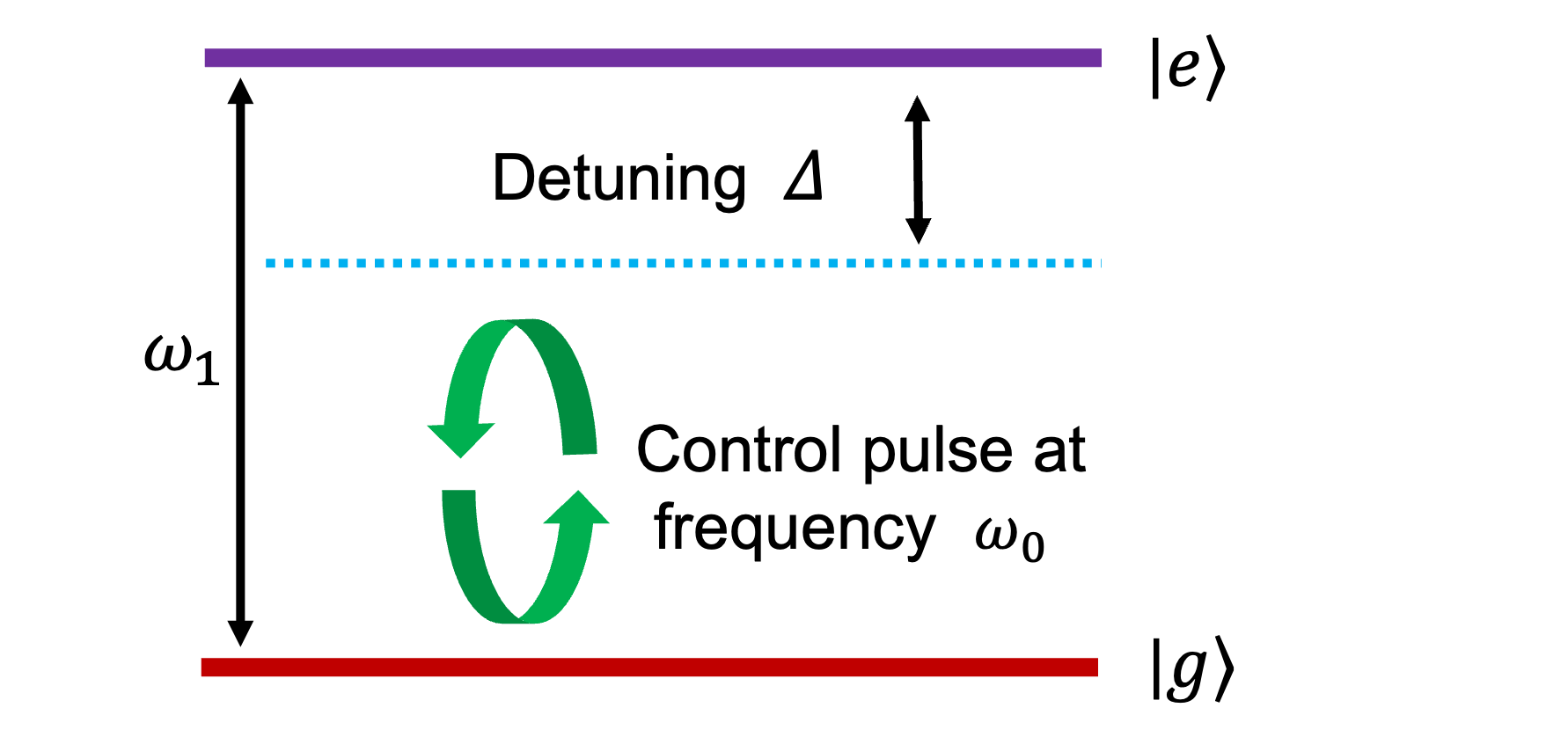}
    %\vspace{-0.5in}
    \caption{ (Color online) A schematic diagram depicting a two-level system (TLS) with its ground state denoted as $\vert g\rangle$ and its excited state as $\vert e\rangle$. The detuning $\Delta$ signifies the offset between the carrier frequency $\omega_0$ and the transition frequency $\omega_1$ of the TLS.
  }
    \label{two-level_sys}
    
\end{figure}

%%%%%%%%%%%%%%%%%%%%%%%%%%%%%%%%%%%%%%%%%%%%%%%%%%%%%%%%%%%%%%%%%%%%%%%%%%%%%%%
We consider a two-level system (TLS) with the ground state $\vert g\rangle$ and excited state $\vert e\rangle$ separated with an energy $\omega_1=\omega_0+\Delta$, with $\Delta$ being a static detuning between the  pulse carrier frequency and the TLS 
%$\omega_0$ from the system's transition frequency $\omega_1$ 
(as shown in Fig~\ref{two-level_sys}). The Hamiltonian of a two-level system interacting with a photonic bath within the rotating wave approximation (RWA), with all energies measured in the frame rotating at the frequency $\omega_0$, is given by ~\cite{cohen1998atom, fotso2017absorption}

\begin{align}
\nonumber
H_0&= \sum_k \omega_k a_k^\dagger a_k + \frac{\Delta}{2} \sigma_z - i \sum_k g_k (a_k^\dagger \sigma_- - a_k \sigma_+), \\ 
%&\quad + \frac{\Omega_x(t)}{2} (\sigma_+ + \sigma_-),
\end{align}
here $a_k$ ($a_k^{\dag}$) represents the annihilation (creation) operator for the $k$-th photon mode, $\omega_k$ is the frequency of mode $k$, $g_k$ denotes the $k$-th photon mode coupling strength with the system of consideration; and  $\sigma_z=\vert e\rangle\ \langle e\vert -\vert g \rangle \langle g\vert $ is the $z-$axis Pauli matrix. 
The TLS coupled to a photon bath is driven by a control laser field, with the control pulse having a time-dependent Rabi frequency $\Omega_x(t)$, applied at appropriate times. The corresponding Hamiltonian of such a driven system is given as:
\begin{align}
\nonumber
H = H_0 + \frac{\Omega_x(t)}{2} (\sigma_+ + \sigma_-),
\end{align}
with $\sigma^{+}=\vert e\rangle\langle g\vert$, and $\sigma^{-}=\vert g\rangle\langle e\vert$ being the Pauli operators for the TLS. 

In the absence of control ($\Omega_x(t)\equiv0$), the system exhibits spontaneous decay and the corresponding emission rate is $\Gamma=2\pi\int g_k^2\delta(\omega_k-\Delta) dk$; the energy and time units are normalized by setting $\Gamma=2$, and we have that the corresponding spontaneous emission line has a simple Lorentzian shape, with half-width equal to $1$. For the controlled case, a square-shaped pulse can be considered for $\Omega_x(t)=\Omega$ during the pulse's active period and remains zero otherwise. In a continuously driven system, the Rabi frequency remains constant over time $\Omega_x(t)=\Omega$.  
%here $\Delta=\omega_1-\omega_0$ is a static detuning of the  pulse carrier frequency $\omega_0$ from the system's transition frequency $\omega_1$. 

%\Omega_x(t)$ is the time-dependent Rabi frequency of the control pulse. 

%{\color{red} $\Omega_x(t)=\Omega$ is the Rabi frequency of the external control pulse. PLEASE ADD more here and how we apply pi pulses}
%\sout{The operators $\sigma_z$, $\sigma^{+} (\vert e\rangle\langle g\vert)$, and $\sigma^{-} (\vert g\rangle\langle e\vert)$ correspond to the $z$ Pauli matrix, the raising and the lowering operators, respectively, for a two-level quantum system. These operators act on the states associated with the system's excited ($\vert e\rangle$) and ground ($\vert g\rangle$) states. Additionally, $a_k$ ($a_k^{\dag}$) represents the annihilation (creation) operator for the $k$-th photon moand $g_k$ denotes its coupling strength with the system of consideration, and $\omega_k$ is the frequency of  mode $k$.}

Starting from the initial state of the density-matrix operator $\rho(t=0)=\rho_{ee}(0)\vert e\rangle\langle e\vert+ \rho_{gg}(0)\vert g\rangle\langle g\vert+\rho_{eg}(0)\vert e\rangle\langle g\vert+\rho_{ge}(0)\vert g\rangle\langle e\vert $, we study the time-dependent dynamics of the TLS by solving the master equations (within the Markovian regime, employing the RWA) of the form ~\cite{cohen1998atom, fotso2016suppressing, fotso2017absorption}:

%\sout {We have the initial state $\rho=\rho_{ee}(0)\vert 1\rangle\langle 1\vert+ \rho_{gg}(0)\vert 0\rangle\langle 0\vert+\rho_{eg}(0)\vert 1\rangle\langle 0\vert+\rho_{ge}(0)\vert 0\rangle\langle 1\vert $ at $t=0$. The behaviour of this two-level system within the Markovian regime, employing the rotating-wave approximation, is described by the master equation~\cite{cohen1998atom}}

\begin{align}
\nonumber
\dot{\rho}_{ee} &= i\frac{\Omega_x(t)}{2}(\rho_{eg}-\rho_{ge})-\Gamma \rho_{ee} \\\nonumber
\dot{\rho}_{gg} &= -i\frac{\Omega_x(t)}{2}(\rho_{eg}-\rho_{ge})+\Gamma \rho_{ee} \\ \nonumber
\dot{\rho}_{ge} &= (i\Delta-\frac{\Gamma}{2})\rho_{ge}-i\frac{\Omega_x(t)}{2}(\rho_{ee}-\rho_{gg}) \\ 
\dot{\rho}_{eg} &= (-i\Delta-\frac{\Gamma}{2})\rho_{ge}+i\frac{\Omega_x(t)}{2}(\rho_{ee}-\rho_{gg}).
\end{align}
%where $\Gamma$ is the decay rate representing the coupling to the radiation bath. %and $\Omega_x (t)$ is the time-dependent driving control pulses coupling the external field with the TLS. In the absence of control ($\Omega_x(t)\equiv0$), the system exhibits spontaneous decay, and the corresponding emission rate is $\Gamma=2\pi\int g_k^2\delta(\omega_k-\Delta) dk$; the energy and time units are normalised and sets $\Gamma=2$, and we have the corresponding spontaneous emission line has a simple Lorentzian shape $\Gamma=1/(\omega^2 + 1)$, with half-width equal to $1$.

The assumption of a strong and brief pulse driving $(\Omega \gg \Delta, \Gamma)$ is adopted, enabling us to treat the pulses as instantaneous. For instance, the application of $\pi_x$ pulses inverts the population of excited and ground states and swaps the values of the coherence terms 
$\rho_{eg}$ and $\rho_{ge}$ as
\begin{equation}
    \rho(n\tau+0)=\sigma_x \rho(n\tau-0)\sigma_x,
\end{equation}
where $\rho(n\tau-0)$ and $\rho(n\tau+0)$ are density matrices immediately before and after the pulse application; $\tau$ is the period of the pulse sequence, $n$ is an integer, and $\sigma_x=\vert g\rangle \langle e \vert +\vert e\rangle \langle g \vert$, i.e., the pulses interchange $\rho_{ee}$ with $\rho_{gg}$, and $\rho_{eg}$ with $\rho_{ge}$.

The evolution of the TLS in the presence of a driving field represents a quantum battery-charger problem. Here, the `battery' discharge, represented by the relaxation from the excited state to the ground state of the TLS, occurs through its interaction with the photonic bath. At the same time, the driving field assumes the role of the charger, supplying energy to the system. The charging-discharging processes for different initial states of TLS for continuously driven and instantaneous pulse-driven scenarios (charging protocols) are examined in the following section.

%%%%%%%%%%%%%%%%%%%%%%%%%%%%%%%%%%%%%%%%%%%%%%%%%%%%%%%%%%%%%%%%%%%%
\section{Method: Ergotropy as a measure of work extraction}\label{ergotropy}
%%%%%%%%%%%%%%%%%%%%%%%%%%%%%%%%%%%%%%%%%%%%%%%%%%%%%%%%%%%%%%%%%%%%%
Ergotropy is the maximum amount of extractable work from a quantum battery using optimal cyclic transformation~\cite{allahverdyan2004maximal, francica2020quantum, ccakmak2020ergotropy,shi2022entanglement}. 
A general Hamiltonian of the quantum battery-charger model can be written as
\begin{equation}
\mathscr{H}(t) = \mathscr{H}_\mathcal{B} + \mathscr{H}_\mathcal{C} + \mathscr{E}(t) ,
\end{equation}
where battery and charger parts are characterised by local Hamiltonians $\mathscr{H}_\mathcal{B}$ and $\mathscr{H}_\mathcal{C}$, respectively. $\mathscr{E}(t)$ is the charging operator that encompasses all terms responsible for regulating energy input, including interactions between the battery and the charger or any external driving fields.
%Initially,  at time $t=0$, the system is prepared in a product state $\rho_{\mathcal{B}\mathcal{C}}(0)$ with the battery being the ground state $\vert \epsilon_0\rangle$ of $\mathscr{H}_\mathcal{B}$. We then suddenly turn on $\mathcal{V}(t)$ and aim to inject as much energy as possible into the battery for a finite time interval $t \in [0, T]$. Such a time interval $T$ is called the charging time. The evolving state is given by $\rho_{\mathcal{B}\mathcal{C}}(t)=U^\dag(t)\rho_{\mathcal{B}\mathcal{C}}U(t)$ with $U(t)=\mathcal{T}\textrm{exp}[-i \int_0^{t}H(t) dt]$. 

Initially, at $t=0$, the state of the battery-charger system is  $\rho_{\mathcal{B C}}(0)$, and its evolved state at time $t$ is denoted by $\rho_{\mathcal{B C}}(t)$. The evolved state of the battery is obtained as $\rho_{\mathcal{B}}(t)=\textrm{Tr}_\mathcal{C}[\rho_{\mathcal{B}\mathcal{C}}(t)]$. The maximum amount of total extractable work (ergotropy) from a state $\rho_{\mathcal{B}}$ can be calculated using the optimal unitary cyclic transformation~\cite{allahverdyan2004maximal}
\begin{equation}    
\textrm{W}(\rho_{\mathcal{B}},\mathscr{H}_\mathcal{B}) = \max_{U \in \mathcal{U}_{\text{cf}}} \left[ E\left( \rho_{\mathcal{B}}(t) \right) - E\left( U\rho_{\mathcal{B}}(t)U^\dagger \right) \right].
\end{equation}
%A close expression for
The ergotropy can be calculated by identifying the passive state $\Tilde{\rho}_\mathcal{B}$ with zero ergotropy, as follows
\begin{equation}
   \textrm{W}(\rho_{\mathcal{B}},\mathscr{H}_\mathcal{B})=E(\rho_{\mathcal{B}})-E(\Tilde{\rho}_\mathcal{B}),
   \label{erg}
\end{equation}
where $E(\rho_\mathcal{B})=\textrm{Tr}(\rho_\mathcal{B} \mathscr{H}_\mathcal{B})$ is the  average energy of the state $\rho_\mathcal{B}$.
$\Tilde{\rho}_\mathcal{B}=\sum_n r_n\vert \epsilon_n\rangle\langle\epsilon_n\vert$ with $\rho_\mathcal{B}=\sum_n r_n\vert r_n\rangle\langle r_n\vert$, $\mathscr{H}_\mathcal{B}=\sum_n \epsilon_n\vert\epsilon_n\rangle\langle\epsilon_n\vert$ with eigenvalues of $\rho_\mathcal{B}$ and $\mathscr{H}_\mathcal{B}$ respectively, $r_0\geq r_1 \geq....r_n$ and $\epsilon_0\leq\epsilon_1\leq....\epsilon_1$, and $E(\Tilde{\rho}_\mathcal{B})$ given by $E(\Tilde{\rho}_\mathcal{B})=\sum_n r_n\epsilon_n$.

It has been shown in Refs.~\cite{francica2020quantum,ccakmak2020ergotropy} that the quantum ergotropy of the battery can be divided into two fundamentally separate components: $\textrm{W}_{\textrm{IC}}$, the incoherent ergotropy and $\textrm{W}_{\textrm{C}}$, the coherent component of the ergotropy so that

\begin{equation}
\textrm{W}(\rho_{\mathcal{B}})=\textrm{W}_{\textrm{IC}}(\rho_{\mathcal{B}})+\textrm{W}_{\textrm{C}}(\rho_{\mathcal{B}}). 
\end{equation}
The incoherent ergotropy $\textrm{W}_{\textrm{IC}} (\rho_{\mathcal{B}})$ presents the maximum work that can be extracted from the $\rho_{\mathcal{B}}$ without changing its coherence and is given as
\begin{equation}
    \textrm{W}_{\textrm{IC}}(\rho_{\mathcal{B}})=\textrm{Tr}[(\rho_{\mathcal{B}}-\pi)\mathscr{H}_\mathcal{B}],
\end{equation}
here $\pi$ is the coherence-invariant state of $\rho_{\mathcal{B}}$ with a lesser average energy such that 

\begin{equation}
    \textrm{Tr}[\pi \mathscr{H}_\mathcal{B}]=\min_{U_i\in U} \textrm{Tr}[U_i \rho U^{\dag}_{i} \mathscr{H}_\mathcal{B}],
\end{equation}
where $U$ is the set of unitary operators on  $\rho_{\mathcal{B}}$ that preserves the coherence but lessens the average energy.
 
Another approach to determine the incoherent contribution to ergotropy ~\cite{francica2020quantum} is to define $\textrm{W}_{\textrm{IC}}$ as the utmost work retrievable from $\rho_{\mathcal{B}}$ after eliminating all its coherences using the dephasing map $\mathcal{M}_{d}$ such that 
\begin{equation}
    \textrm{W}_{\textrm{IC}}(\rho_{\mathcal{B}})\equiv \textrm{W}(\mathcal{M}_d(\rho_{\mathcal{B}}))=\textrm{W}(\rho_{\mathcal{B}}^{d})= \textrm{Tr}(\mathscr{H}_\mathcal{B}(\rho_{\mathcal{B}}^{d}-\rho_{\mathcal{B} p}^{d})),
    \label{ergo_incoherent}
\end{equation}
here $\rho_{\mathcal{B}}^{d}$ has the same population as $\rho_{\mathcal{B}}$ but zero coherence. $\rho_{\mathcal{B} p}^{d}$ is the passive state corresponding to $\rho_{\mathcal{B}}^{d}$, and is obtained from $\rho_{\mathcal{B}}^{d}$ after rearranging the population in the descending order. 
% \sout{The calculation of incoherent ergotropy gives the coherent contribution as} 
The coherent contribution to entropy is obtained as follows
\begin{equation}
    \textrm{W}_\textrm{C} (\rho_{\mathcal{B}})=\textrm{W}(\rho_{\mathcal{B}})-\textrm{W}_{\textrm{IC}}(\rho_{\mathcal{B}}).
\end{equation}

\section{Results}\label{erg_TLS}
This section presents our results for the coherent and incoherent contributions to the total ergotropy obtained for our TLS model. We analyze the ergotropy dynamics for different initial states under both continuous and pulse-driven charging protocols. Specifically, we consider two distinct initial states of the TLS during the charging-discharging cycles: (i) the system is initially in the excited state, $\vert e\rangle$, which possesses maximum initial energy, and (ii) the system is initially in the maximally coherent state, $\frac{1}{\sqrt{2}}(\vert e\rangle + \vert g\rangle)$, characterized by maximum coherence. These states serve as representative cases for two important classes of TLS states. % The ergotropy is employed as a measure of the extractable work from the TLS and 
We estimate both the coherent and incoherent contributions to the total ergotropy for each initial state. To achieve this, we explore the dynamics of work extraction under continuous and periodically driven charging strategies, providing insight into the role of initial conditions and how the control fields influence the efficiency of energy extraction.

\subsection{Work extraction from a continuously driven TLS}

\begin{figure}[!t] 
\centering
\includegraphics[ width=0.5\textwidth] {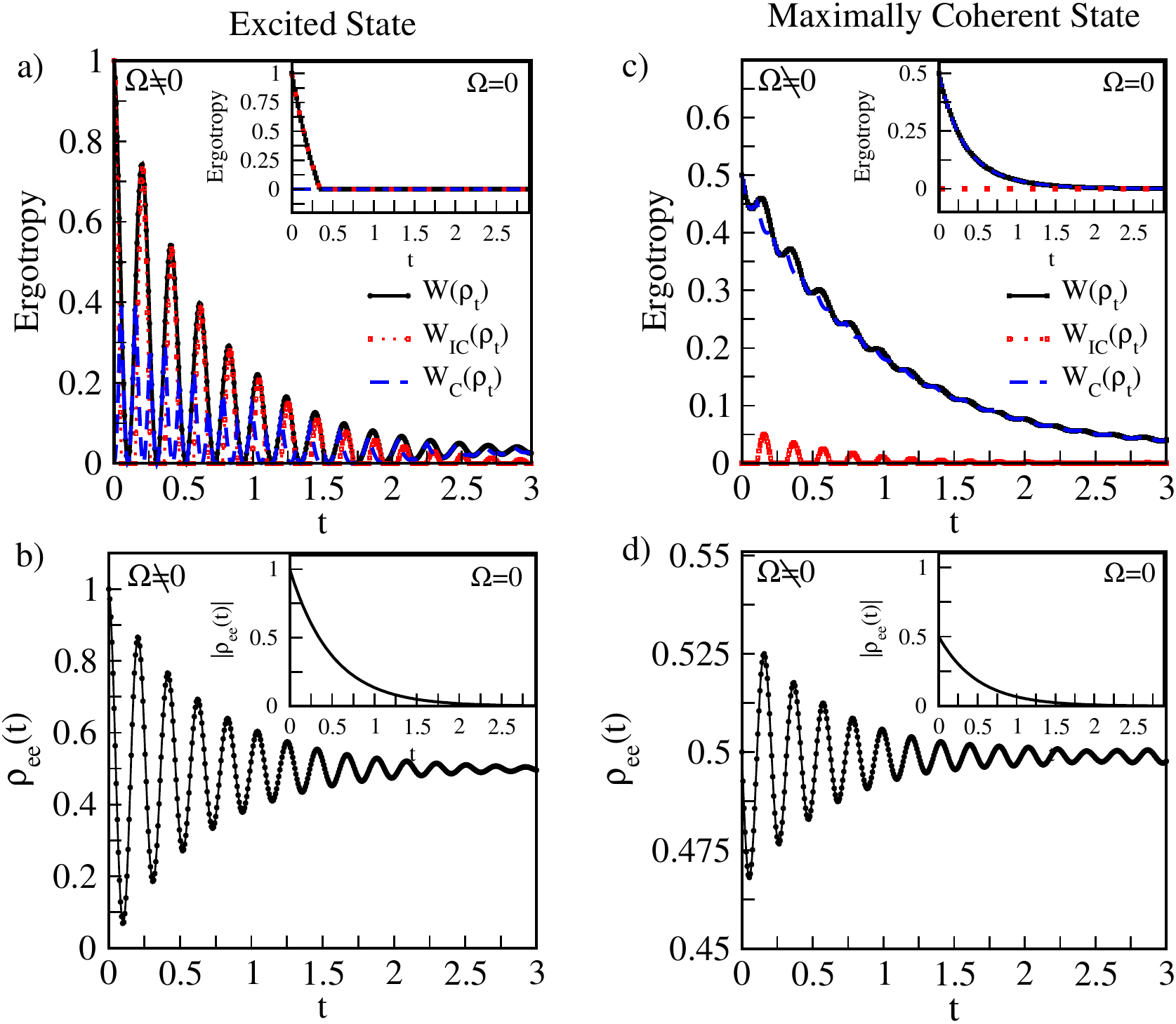}
%{erg_CD_es_cs.eps}
% \includegraphics[height=65mm,width=1\columnwidth]{ergDes.png}
\caption{ (Color online) Variation of total ergotropy ($\textrm{W}$), incoherent and coherent ergotropy ($\textrm{W}_\textrm{{IC}}$, $\textrm{W}_\textrm{C}$), and population dynamics ($\rho_{ee}$) for two initial states: the excited state (left panels: a and b) and the maximally coherent state (right panels: c and d). The system interacts with a photonic bath driven by a field with Rabi frequency $\Omega = 30$ in the resonance regime ($\Delta = 0$). Insets display the corresponding quantities without the external control field ($\Omega = 0$).}
\label{erg_D_es}
\end{figure}

We begin our analysis by exploring the behaviour of a continuously driven quantum emitter \sout{TLS} under finite Rabi oscillation ($\Omega = 30$) in the resonance regime ($\Delta = 0$). Figure~\ref{erg_D_es} displays the time evolution of total ergotropy ($\textrm{W}$), coherent ergotropy ($\textrm{W}_\textrm{C}$), incoherent ergotropy ($\textrm{W}_\textrm{{IC}}$), and the excited state population ($\rho_{ee}$). The left panel illustrates these dynamics for the fully excited initial state ($\vert e\rangle$), while the right panel shows the behaviour for the maximally coherent initial state ($\frac{1}{\sqrt{2}}(\vert e\rangle + \vert g\rangle)$). The insets show the corresponding results for the non-driven case ($\Omega = 0$).
%To highlight the influence of the driving field on the system's evolution, the insets present the corresponding results for the undriven case ($\Omega = 0$). 

When the control field is applied ($\Omega\neq 0$), energy is injected into the  TLS, akin to a charging process in a quantum battery. Dissipation and decoherence of the system result from its interaction with the photonic bath, leading to energy discharge from the system. The variation in work extraction of a continuously driven TLS illustrates the charging-discharging process of a quantum battery. For the excited initial state shown in Figs.~\ref{erg_D_es} (a) and \ref{erg_D_es} (b), such a charging-discharging cycle of the quantum battery follows the population dynamics, this is seen by the oscillating behaviour of the ergotropy ($\textrm{W}$) which resembles the dynamics of the excited state population $\rho_{ee}$. As seen from   Fig.~\ref{erg_D_es} (a), both coherent and incoherent components contribute significantly to the total ergotropy. The incoherent part $\textrm{W}_\textrm{{IC}}$  shows oscillatory behaviour, vanishes for some time, as its excited-state population decreases below the steady-state population, then revives back, and gradually diminishes as the system approaches the steady state.  The coherent ergotropy $\textrm{W}_\textrm{C}$ initially exhibits large amplitudes, but these amplitudes diminish as the state approaches the steady state, with fluctuations occurring around lower values.

Similarly, in Fig.~\ref{erg_D_es} (c) and \ref{erg_D_es} (d), we show the ergotropy ($\textrm{W}$) and population dynamics ($\rho_{ee}$) for the maximally coherent initial state. As can be seen from Fig.~\ref{erg_D_es} (c), there is a little incoherent contribution ($\textrm{W}_\textrm{{IC}}$) to the total work extraction in the early stages of the dynamics, and the incoherent ergotropy vanishes as the excited state population decreases below its steady-state, and revives back, $\textrm{W}_\textrm{{IC}}$ later vanishes as the system approaches the steady state. In such a maximally coherent state, coherence is the main contributor to the total ergotropy, evident by the dominant contribution of $\textrm{W}_\textrm{C}$ to the total ergotropy. We also observe that unlike in the excited initial state, for the coherent case, the oscillatory nature of ergotropy $\textrm{W}$ is less pronounced, which is also reflected in the population dynamics (Fig.~\ref{erg_D_es} (d)) oscillating with smaller amplitude.

%fluctuating between lower values.

To highlight the influence of the driving field on the system's evolution, the insets present the corresponding results for the non-driven case ($\Omega = 0$). In the absence of an external control field, the excited state population evolves as $\rho_{ee}(0) e^{-\Gamma t}$. For our initial excited and coherent states, $\rho_{ee}(0)$ takes the values $1$ and $\frac{1}{2}$, respectively, which decay exponentially with time, as shown in Fig.~\ref{erg_D_es} (b-d) insets. Unlike the driven case with the oscillating behaviour, the work extraction dynamics of an initial TLS state without any control field resembles the discharging process of a quantum battery without any intermediate charging methods (Fig.~\ref{erg_D_es} (a) and (c) insets). For the fully excited state, the coherent ergotropy $\textrm{W}_\textrm{{C}}$ contribution to the total ergotropy $\textrm{W}$ is zero. In contrast, for the maximally coherent initial state, the incoherent part $\textrm{W}_\textrm{{IC}}$ does not contribute to the total ergotropy.

\begin{figure}[!t]
\centering
\includegraphics[width=0.5\textwidth]{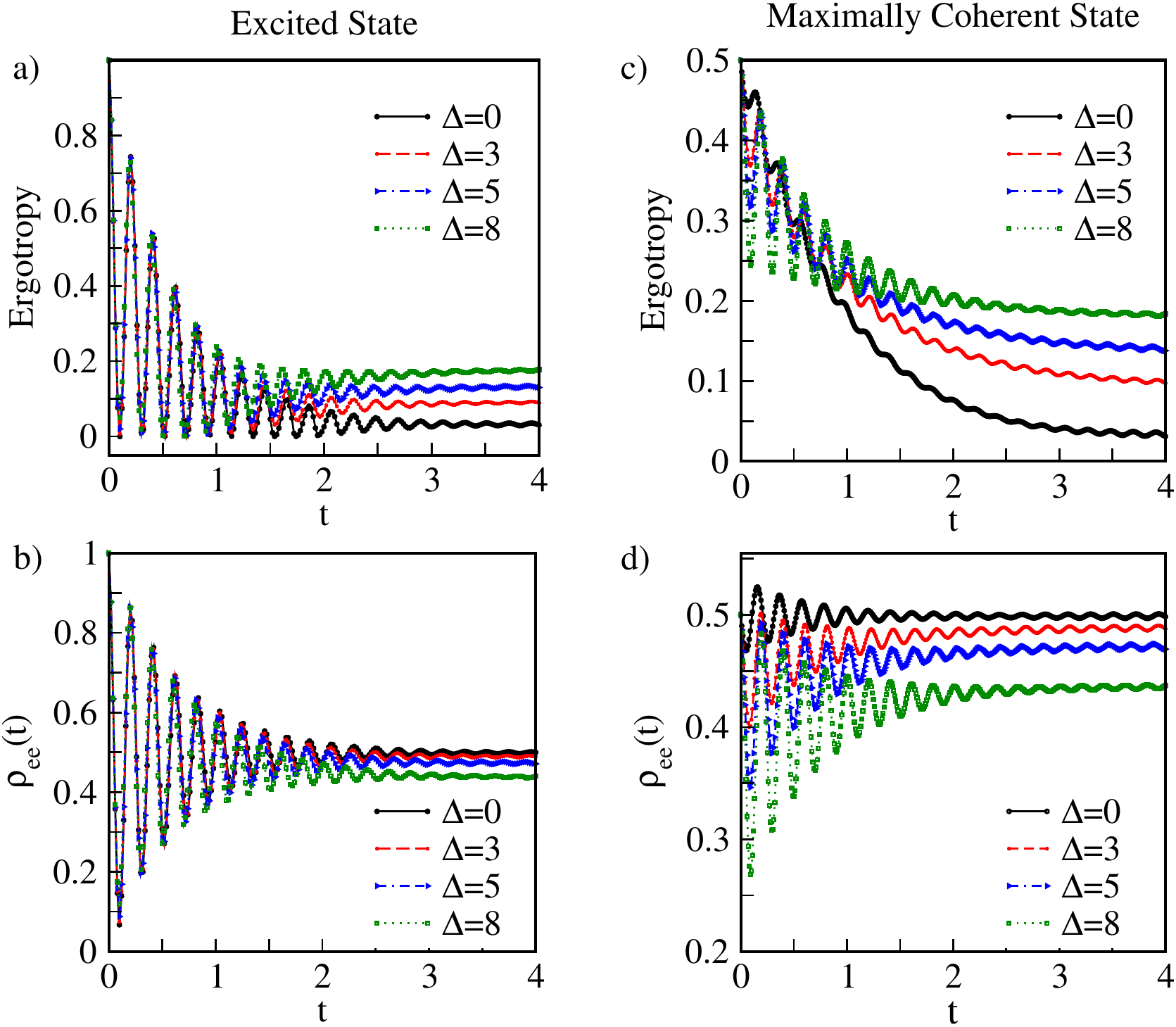} %{erg_CD_Dvs.eps}
\caption{ (Color online) Time-dependent dynamics of the total ergotropy ($\textrm{W}$) and the excited state population ($\rho_{ee}$) for different initial conditions and values of detuning ($\Delta$). The left panels (a and b) show the dynamics for an initial excited state, while the right panels (c and d) represent the dynamics for a maximally coherent initial state. The system is driven by a field with Rabi frequency $\Omega = 30$.}
\label{erg_D_Dvs}
\end{figure}
Next, we consider the influence of detuning $\Delta$ on the variation of ergotropy $\textrm{W}$. Our results for the total ergotropy $\textrm{W}$ and the corresponding excited state population dynamics $\rho_{ee}$ for different values of detuning $\Delta=0, 3, 5, 8$  for the fully excited initial state (left panel) and maximally coherent initial states (right panel) are shown in Fig.~\ref{erg_D_Dvs}. 
%The ergotropy variation and corresponding population dynamics for the excited state for different values of detuning $\Delta$ are depicted in Figs.~\ref{erg_D_Dvs} (a) and \ref{erg_D_Dvs} (b), respectively. Similarly, we have them for maximally coherent state in~Figs.~\ref{erg_D_Dvs} (c) and \ref{erg_D_Dvs} (d).
We show that ergotropy from the continuously driven TLS increases as the value of  $\Delta$ increases for both excited and coherent states (Figs.~\ref{erg_D_Dvs} (a) and \ref{erg_D_Dvs} (c)) as the system approaches steady-state dynamics. For the system initially in the excited state, from Fig.~\ref{erg_D_Dvs} (a), it is evident that the ergotropy as a function of time increases as the detuning increases. However, this is different in the case of the maximally coherent state. For the maximally coherent state (Fig.~\ref{erg_D_Dvs} (a), the dependence of ergotropy on detuning follows a reverse order at the early stage of the dynamics due to the difference in the (in)coherent ergotropy contribution to the total ergotropy. However, this changes as the system approaches steady state dynamics, where the ergotropy increases as $\Delta$ increases. In the left panel (a), the system initially in the excited state exhibits more pronounced oscillations, which diminish in amplitude as the steady state approaches. For the system initially in the coherent state (right panel (c)), the ergotropy in the early stage shows oscillatory behaviour for higher values of detuning $\Delta$. As the system approaches its steady state, the oscillation of ergotropy decreases in amplitude, revealing a more pronounced dependence on $\Delta$. 
Panels (b) and (d) show the corresponding population dynamics of the excited state, where similar oscillations are observed for both initial states. However, the population dynamics of the maximally coherent state (panel (d)) fluctuates between lower values, and the oscillations are less prominent compared to those of the excited state (panel (b)). The effect of detuning $\Delta$ is evident in all panels, with higher detuning ($\Delta$) leading to a slower decay in ergotropy and a shift in population dynamics. Overall, the figure highlights that detuning $\Delta$ can be used to manipulate the ergotropy one can extract from the TLS battery over time.

\subsection{Work extraction from a periodically-driven TLS}

\begin{figure}[!tb]
    \centering
    \includegraphics[width=0.5\textwidth]{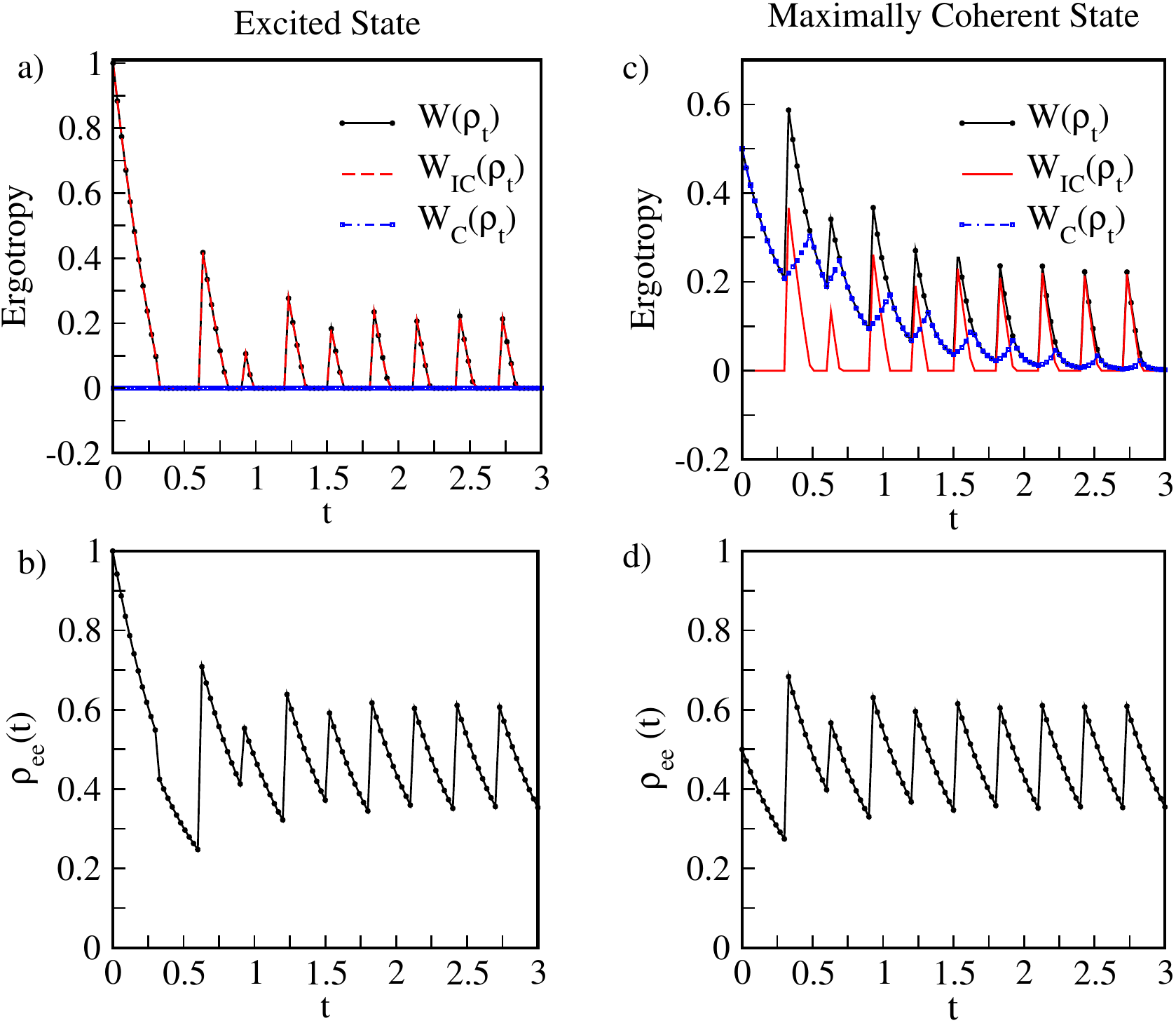}
    \caption{(Color online) 
    Variation of total ergotropy $\textrm{W}$, incoherent $\textrm{W}_{\textrm{IC}}$, and coherent ergotropy $\textrm{W}_\textrm{C}$, along with the population dynamics ($\rho_{ee}$) for the initially excited state (left panels) and the maximally coherent state (right panels). The system interacts with a photonic bath driven by instantaneous $\pi_x$ pulses with a $\tau = 0.3$ period. Number of pulses $N_{pulse} = 10$, $\Delta = 0$.}
    %Variation of total ergotropy, (in)coherent ergotropy, and the population dynamics of the excited state (Left panel: a and b) and the maximally coherent state (Right panel: c and d), interacting with a photonic bath driven by $\pi_x$ pulses with a period of $\tau=0.3$, after 10 pulses, in the resonance regime ($\Delta=0$).}
    \label{erg_PD_es}
\end{figure}

\begin{figure}[!tb]
    \centering
    \includegraphics[width=0.5\textwidth]{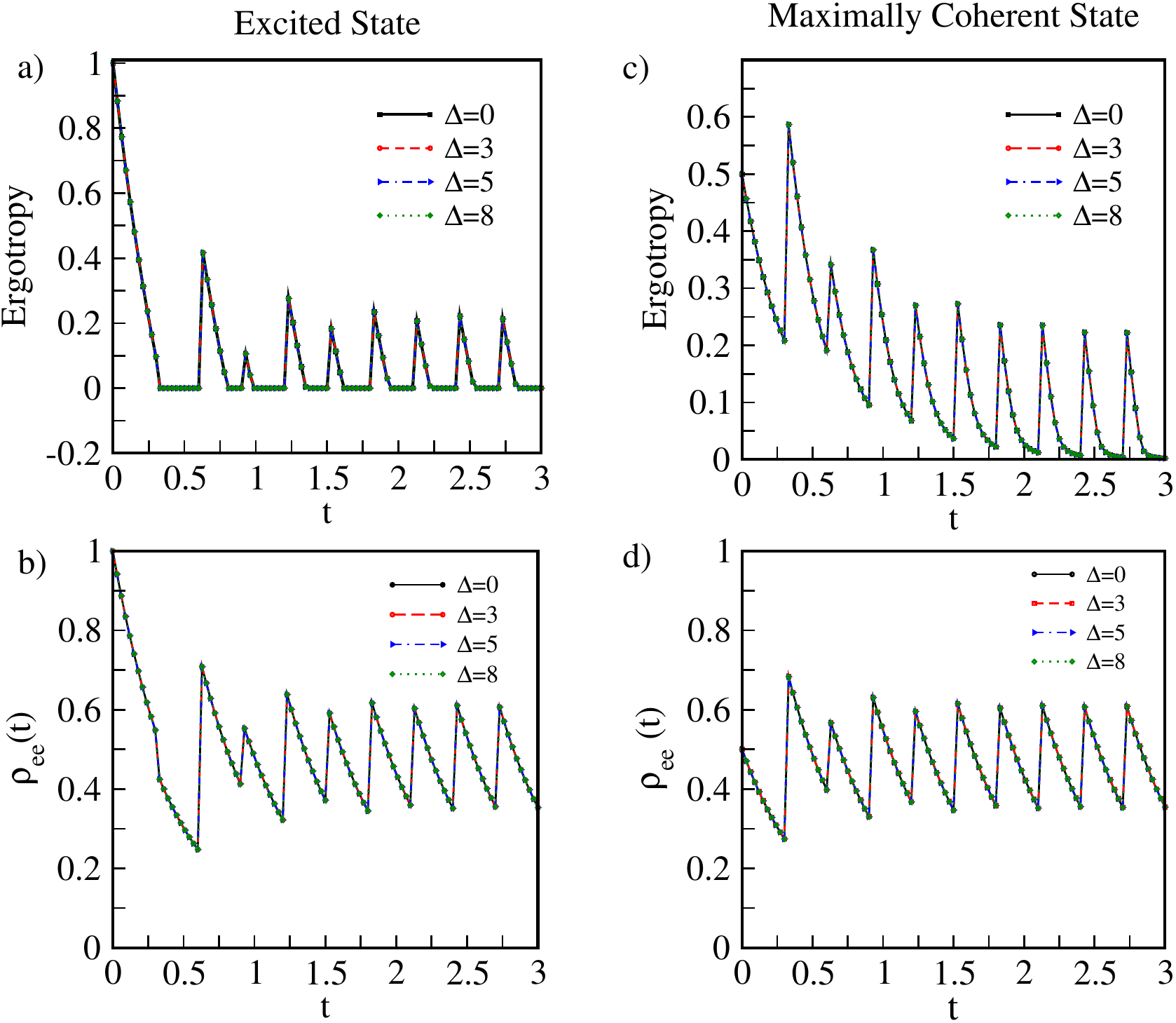}
    \caption{ (Color online) Variation of total ergotropy $\textrm{W}$ and the population dynamics $\rho_{ee}$ of the initially excited state (left panels) and the maximally coherent state (right panels), interacting with a photonic bath driven by instantaneous $\pi_x$ pulses with a period of $\tau=0.3$. Number of pulses $N_{pulse} = 10$, values of detuning $\Delta=0, 3, 5, 8$.}
    \label{erg_PD_es_cs_Dvs}
\end{figure}

In this section, we analyze the work extraction from a quantum emitter driven by a periodic sequence of $\pi_x$ pulses, which act as a control mechanism applied instantaneously—similar to the instantaneous charging of a quantum battery. Following each pulse, the battery discharges according to the Markovian dynamics. Each $\pi_x$ pulse swaps the populations of the excited and ground states while exchanging their coherence values, repeating at intervals of $\tau$. 

Fig.~\ref{erg_PD_es} shows the discharge dynamics of the quantum battery for two initial states in the resonance regime ($\Delta = 0$), highlighting how the ergotropy and population dynamics evolve under the influence of a periodic sequence of $\pi_x$ pulses and emphasizing the differences between the initial state excited or maximally coherent. For the initial excited state (left panel, Fig.~\ref{erg_PD_es} (a) and (b), the $\pi_x$ pulses applied periodically at a time interval $\tau = 0.3$ result in the total ergotropy decaying monotonically for one inter-pulse time interval and then vanishing for the following one.  Here, the total ergotropy $\textrm{W}$ consists solely of the incoherent part $\textrm{W}_\textrm{{IC}}$, with no contribution from coherence, as shown by the zero value of $\textrm{W}_\textrm{C}$. The population dynamics of the excited state ($\rho_{ee}$) exhibit a sharp, step-like decrease with each pulse, inverting the population between the excited and ground states. Total ergotropy vanishes as the excited-state population decreases below the steady-state population.  For the maximally coherent initial state (Fig.~\ref{erg_PD_es} (c) and (d), both the coherent $\textrm{W}_\textrm{C}$ and incoherent $\textrm{W}_\textrm{IC}$ components contribute to the total ergotropy $\textrm{W}$. The coherent ergotropy plays a significant role alongside the incoherent ergotropy, which diminishes as the excited state population decreases below the steady state population, resulting in oscillatory behaviour in the total extractable work. The interplay between these contributions is evident from the population dynamics in (Fig.~\ref{erg_PD_es} d)), where the periodic $\pi_x$ pulses drive oscillations between the excited and ground states. The dynamics of the incoherent ergotropy are closely tied to these population oscillations, offering insight into how the population inversion impacts the extractable work from the system.

In Fig.~\ref{erg_PD_es_cs_Dvs}, we investigate the effect of detuning on work extraction in a TLS driven by a periodic sequence of instantaneous $\pi_x$ pulses. Unlike the case for a continuously driven system in the previous section, the detuning does not significantly alter the charging-discharging behaviour for either initial state. Each $\pi_x$ pulse injects energy into the system instantaneously, acting like a rapid charge to a quantum battery, followed by a discharge phase before the next pulse is applied. We also examined the case of a finite-duration pulse (results not shown) and observed behaviour similar to that of a continuously driven system. When a finite-duration $\pi$ pulse is applied for time $\frac{\pi}{\Omega}$, the effect of detuning on ergotropy is noticeable for smaller Rabi frequencies, especially in the near-resonant regime. 

\section{Conclusion}\label{conclusion}
In this work, we explored how a controlled two-level quantum system can serve as a model for a quantum battery when interacting with a photonic bath. We systematically analyzed work extraction under continuous  and periodic pulse field protocols, using ergotropy to quantify the extractable work. Our investigation considered two initial states of the TLS — a fully excited and a maximally coherent initial state — to explore their respective influences on the charging dynamics. We also considered the coherent and incoherent contributions to the ergotropy,  finding that the nature of these contributions depends on the initial state.

For a continuously driven quantum emitter, we have shown that total ergotropy increases with detuning, particularly in the near-resonant regime, where higher detuning leads to greater extractable work. This trend is observed in the population dynamics of the system's excited state. % and is consistent with a higher driving Rabi frequency.

In the case of a continuously driven TLS, the contributions of coherent and incoherent ergotropy to the total extractable work show distinct behaviours with different initial states. The incoherent ergotropy ($\textrm{W}_\textrm{{IC}}$) primarily arises from the population dynamics of the TLS, and its variation follows the oscillations and decay patterns of the excited state population. The incoherent ergotropy vanishes as the excited state population decreases below the steady state population. For the system initially in the excited state, the incoherent ergotropy contribution to $\textrm{W}$ is more significant at the beginning of the dynamics but diminishes over time as the system approaches steady-state behaviour. In contrast, for the maximally coherent state, the contribution of incoherent ergotropy to the ergotropy is less significant.  The coherent ergotropy ($\textrm{W}_\textrm{{C}}$), on the other hand, is linked to the coherence of the TLS states. As the detuning increases, particularly in the near-resonant regime, $\textrm{W}_\textrm{{C}}$ contributes significantly to the total ergotropy, resulting in a more significant amount of work that can be extracted from the system.  The coherent contribution to the ergotropy varies with the initial states. During the initial phase of evolution, the large amplitude oscillations in the excited state population result in the incoherent contribution to ergotropy, surpassing the coherent contribution for some time. This trend changes as the system approaches its steady state, where coherence significantly contributes to the ergotropy. In the case of the system initially in the maximally coherent state, coherence contributes considerably to the ergotropy throughout the evolution, and incoherent ergotropy is less critical.

In the case of periodic $\pi_x$ pulse-driven charging, our results showed that the detuning plays a negligible role in the ergotropy dynamics, which contrasts with the continuous driving scenario. Here, the work extraction is primarily influenced by the inter-pulse delay $\tau$, where longer delays lead to higher peaks in ergotropy. 
%{\color{red}Furthermore, we examined the effect of finite-duration $\pi$ pulses and found that the ergotropy dynamics converge to those of a continuously driven system as the inter-pulse interval is reduced, demonstrating that pulse duration and timing are crucial parameters for optimal control.}\newline
Overall, our findings underscore the importance of both control strategy and initial state in maximizing the work extraction from a quantum battery. By systematically analyzing how different parameters affect the ergotropy dynamics, this work provides insights into the operation of efficient quantum batteries and highlights the effect of control protocols in optimizing energy storage and retrieval in quantum systems.

\section*{Acknowledgement}
KGP and HT are supported by the National Science Foundation under Grant No. QIS-2328752. HT is supported by the NSF DMR-1944974 grant. HFF is supported by the National Science Foundation under Grant No. PHY-2014023 and QIS-2328752.

\section*{Appendix}
\renewcommand{\theequation}{I.\arabic{equation}}
\setcounter{equation}{0}
\section*{Ergotropy calculation for a single qubit state}
Consider the qubit's state,
\begin{equation}
\rho = \frac{1}{2} \left( \sigma_0 + r_1 \sigma_x + r_2 \sigma_y + r_3 \sigma_z \right),
\end{equation}
where $\sigma_{i=1,2,3}$ are the Pauli's spin matrix and $\sigma_{0}$ the identity matrix.
Consider the the Hamiltonian of the form $H=\frac{\omega_0}{2}(\sigma_0+\sigma_z)$ on the qubit state. We calculate the ergotropy of the state using Eq.~\ref{erg} 
\begin{equation}
    \textrm{W}(\rho)= \frac{\omega_0}{2}(r+r_3),
\end{equation}

and $r=\sqrt{r_1^2+r_2^2+r_3^2}$. Using Eq.~\ref{ergo_incoherent}, we calculate the incoherent ergotropy. We have $\rho^{d}=\frac{(1-r_3)}{2}\vert 1\rangle\langle 1\vert+ \frac{(1+r_3)}{2}\vert 0\rangle\langle 0\vert$, $\vert 1\rangle=(1,0)^T$ and $\vert 0\rangle= (0,1)^T$. Eigenvalues of 
$\rho^{d}$ in the decreasing order are $\frac{1-r_3}{2}\geq\frac{1+r_3}{2}$ for $-1\leq r_3\leq 0$, and for $0<r_3\leq 1$, we have  $\frac{1+r_3}{2} \geq \frac{1-r_3}{2} $. We calculate $\rho^{d}_p$ as
$\frac{(1+r_3)}{2}\vert 1\rangle\langle 1\vert+ \frac{(1-r_3)}{2}\vert 0\rangle\langle 0\vert$ and $\frac{(1-r_3)}{2}\vert 1\rangle\langle 1\vert+ \frac{(1+r_3)}{2}\vert 0\rangle\langle 0\vert$ for the intervals $-1\leq r_3\leq 0$ and $0<r_3\leq 1$, respectively.\newline
Incoherent ergotropy for the quantum state $\rho$ are $0$ and $r_3 \omega_0$, for $-1\leq r_3\leq 0$, and $0<r_3\leq 1$, respectively.  We calculate the coherent contribution to the ergotropy for the  two previously considered intervals of $r_3$ as $\frac{\omega_0}{2}(r+r_3)$ and 
$\frac{\omega_0}{2}(r-r_3)$.

\bibliography{main.bib}% Produces the bibliography via BibTeX.

\end{document}